\documentclass[prc,preprint,showpacs,showkeys,nofootinbib]{revtex4}%
\usepackage{graphicx}
\usepackage{bm}
\usepackage{epsfig}
\usepackage{amsmath}
\usepackage{amsfonts}
\usepackage{amssymb}%
\begin{document}
\title{Proton-tetraneutron elastic scattering}
\author{B.M. Sherrill and C.A. Bertulani}
\email{sherrill@nscl.msu.edu, bertulani@nscl.msu.edu}
\affiliation{National Superconducting Cyclotron Laboratory, and Department of Physics and
Astronomy, Michigan State University, East Lansing, MI\ 48824-1321, USA}
\date{\today}

\begin{abstract}
We analyze the elastic scattering of protons on a $^{4}$n system.
This was used as part of the detection technique of a recent
experiment \cite{Ma02} to search for the $^{4}$n (tetraneutron) as
a bound particle. We show that it is unlikely that this process
alone could yield the events reported in ref. \cite{Ma02}, unless
the $^{4}$n has an anomalously large backward elastic scattering
amplitude.

\end{abstract}
\maketitle

\draft

\address{National Superconducting Cyclotron Laboratory, and
Department of Physics and Astronomy, Michigan State University, E.
Lansing, MI 48824-1321}

\narrowtext

The possible existence of the tetraneutron has been discussed
theoretically by numerous authors already in the 1960's (see refs.
\cite{Ma02,Tim02,BZ03} and references therein). Numerous
experiments have been performed with negative results
(\cite{Ma02,Tim02,BZ03} and references therein). However, in a
recent experiment \cite{Ma02}, the existence of bound neutron
clusters was studied by fragmentation of intermediate energy (35
MeV/nucleon) $^{14}$Be nuclei. As described in ref. \cite{Ma02},
the data analysis was confined (for technical reasons related to
the detector response) to 11-18 MeV/nucleon. The fragmentation
channel $^{10}\mathrm{Be}+\ ^{4}\mathrm{n}$ was observed and the
$^{4}\mathrm{n}$ system was tentatively described as a bound
tetraneutron system.

Theoretical attempts have failed to find the mechanism \ for the
$^{4}\mathrm{n}$\ binding. Ref. \cite{Tim02} concluded that a strong
four-nucleon force is needed to bind the tetraneutron. This force would
unreasonably bind $^{4}$He by about 100 MeV. Ref. \cite{BZ03} showed that a
model of the tetraneutron based on two dineutron molecules is also not likely
to yield a bound system. Further, ref. \cite{Pie03} showed that if the $^{4}$n
would form a bound system it would indeed look like two widely separated
dineutrons, with rms radii between 7.3 fm and 10.3 fm. However, these
conclusions were based on a significant change of presently adopted NN potentials.

In this brief report we point out a puzzle in the experimental
measurement \cite{Ma02}. We calculate the elastic
proton-tetraneutron cross section and show that it is surprising
that the experimental technique used in ref. \cite{Ma02}\ could
identify a bound  $^{4}$n based on this reaction. In the original
experimental paper\ the $^{4}$n was identified by the recoil
imparted on a proton target. In particular $^{4}$n\ were
identified by an anomalous large backward recoil of the proton. We
claim that, due to the loosely bound character of the tetraneutron
(if it existed) this anomalously large proton recoil is unlikely,
as the elastic cross section for the proton-tetraneutron system
drops many orders of magnitude for large scattering angles. If the
data of ref. \cite{Ma02}\ \ is correct, they indicate an
anomalously large backward angle cross section or that other
processes are contributing.

To calculate the proton-$^{4}$n scattering cross section one needs the
p-$^{4}$n phase-shifts $\delta_{l\pm}$, or an optical potential, both of which
are of course unknown. One can construct one theoretically using several
recipes. For simplicity we will use here the M3Y interaction \cite{Ber77} to
construct $U_{\mathrm{\operatorname{real}}}\left(  r\right)  $. The imaginary
part of the potential is taken as $U_{\mathrm{imag}}\left(  r\right)  =\lambda
U_{\mathrm{\operatorname{real}}}\left(  r\right)  $. We adopt a very
conservative value, $\lambda=1$, for simplicity. The input in this calculation
is the tetraneutron density distribution $\rho\left(  r\right)  $ which is of
course unknown. We assume that if the tetraneutron exists it has a large rms
radius, due to its loosely-bound character. We take a one-body density
distribution with rms radii of 7 fm and 10 fm, respectively \cite{Piecom}.
This simulates the small binding energy of the system.

The result of the calculation is presented in figure 1, for
$E_{p}=14$ MeV and a $^{4}$n-system with rms radii equal to 7 fm
(dashed line) and 10 fm (solid line), respectively.

\begin{figure}[t]
\begin{center}
\includegraphics[
height=3.8994in,
width=3.9176in
]{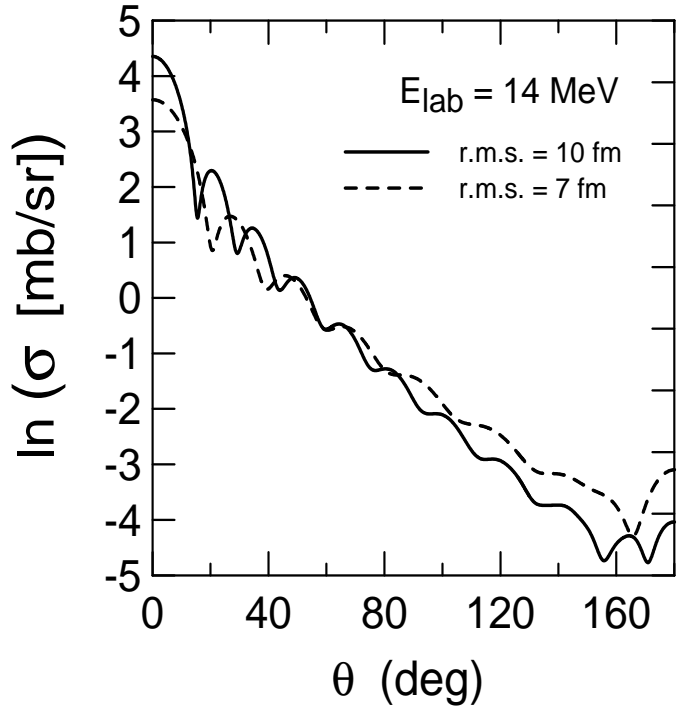}
\end{center}
\caption{Proton-$^{4}$n elastic scattering at $E_{p}=14$ MeV. The
dashed (solid) line assumes a $^{4}$n matter density distribution
with an rms radius
of 7 fm (10 fm).}%
\end{figure}

In figure 2 we show the kinematics of the $^{4}$n-proton
scattering in the laboratory. In particular, we show the proton
energy as a function of its backscattering angle in degrees, in
the kinematical allowed region. In the laboratory, the protons
would have recoil energies in the interval $E_{p}=21$ - 38 MeV,
for scattering angles between 95 - 180 degrees. This is shown by
the solid line in figure 2. The dashed line is the respective
$^{4}$n energy.

For proton recoil energies in the interval $E_{p}/E_{n}=1.4$ - 2.6
we get a total elastic cross section (from figure 1) of about 3.5
$\mu$b, assuming a $^{4}$n size of 7 fm (a $^{4}$n size of 10 fm
yields a 10 times smaller value).

The intensity of $^{14}$Be in the experiment of ref. \cite{Ma02}
is 130 particles per second. Assuming a $^{4}$n production cross
section of 1 mb, as estimated in ref. \cite{Ma02}, as an upper
estimate and a carbon target with 275 mg/cm$^{2}$, we obtain
$1.8\times 10^{-2}$ $^{4}$n/s. The density of protons in the
target can be estimated as $n=10^{24}$/cm$^{2}$. For a cross
section of 3.5 $\mu$b one thus gets a counting rate of 10$^{-9}$
scattered protons per second. The overall (geometry + response of
the detectors) detector efficiency in the experiment is of the
order of 10\%. This yields approximately $10^{-4}$ events per
week. The experiment of ref. \cite{Ma02} claims 6 events in a one
week run. There is a factor of 10$^{4}$ more events than expected
from an optimistic estimate.

The estimates presented above show that the results of the experiment can not
be explained as arising from elastic scattering unless there is an anomalous
large backscattering in the $^{4}%
$n+p system. We think that the $^{4}$n model used in our calculations should
yield an order of magnitude estimate of the distribution of the backscattered
protons. It is well know that the elastic scattering cross section decreases
exponentially with the diffuseness of the system. For a system of size $R$ and
diffuseness $a$, the differential cross sections scales as (see, e.g., ref.
\cite{Fes93}) $d\sigma/d\Omega\sim\left[  J_{1}\left(  qR\right)  /q\right]
^{2}\exp\left(  -qa\right)  $, where $q=2k\sin(\theta/2)$ is the momentum
transfer. Thus, the slope of the cross section in figure 1 is due almost
entirely to the diffuse character of the $^{4}$n system. This behavior is
universal and would not change with a better description of the potential. In
the absence of the Coulomb interaction and other effects, the peak at forward
angles yields the total cross section by means of the optical theorem. Its
magnitude is due to the overall size of the system. Thus, the only parameter
of interest in this case is the diffuseness of the tetraneutron, which
determines the backward/forward cross section ratio.

One possible scenario to explain the events seen in ref. \cite{Ma02} is that
the neutrons do not scatter from the $^{4}$n as a whole, but from a smaller
neutron cluster inside the $^{4}$n system. If the dineutron exists, it might
be less diffuse than the $^{4}$n system. This in consequence could lead to a
larger scattering cross section of the protons towards large angles. However,
this would also lead to scattered protons with smaller energy. Another
possible effect, short-range correlations, which imply large momentum
components inside the $^{4}$n system, are also ruled out as they must be a
small component of the total $^{4}$n wavefunction. We thus conclude that both
of these effects do not help in explaining those events.

\begin{figure}[ptb]
\begin{center}
\includegraphics[
height=4.in, width=4 in ] {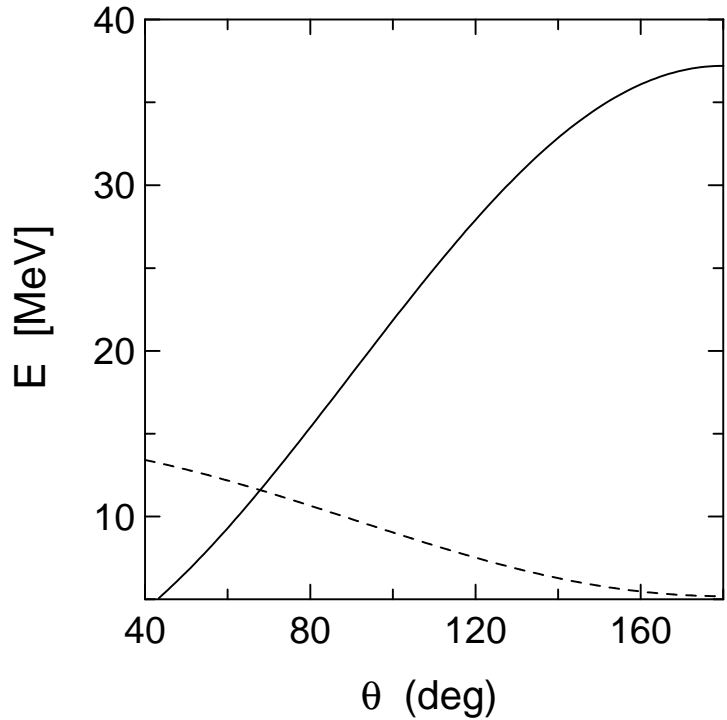}
\end{center}
\caption{ Kinematics of the $^{4}$n-proton scattering in the
laboratory. The solid line shows the proton recoil energies in the
interval $E_{p}=21$ - 38 MeV, for scattering angles between 95 -
180 degrees.  The dashed line is the
respective $^{4}$n energy (per nucleon).}%
\end{figure}

One should also expect that additional effects probably imply an
even smaller cross section for the $^{4}$n (or $^{2}$n)-p elastic
scattering at large angles. Since $^{4}$n is loosely bound there
will be a large absorption from the elastic channel due to the
inelastic events which disrupt the $^{4}$n system. The larger the
scattering angle, the larger the momentum transfer and the smaller
the cross section will be. This effect was not included in our
calculations and would pose an even worse scenario for the
detection of elastic scattered $^{4} $n by proton backscattering.

Our calculations suggest that other processes besides elastic
scattering are required to sustain the bound 4n interpretation of
the events observed in ref. [1]. A possible case would be the
inelastic scattering on protons of very weakly bound systems
\cite{Li11}. However, to produce $\theta_{cm}
> 100^\circ$ recoils, the $^4$n system must interact as a
composite object. Given its weak binding, it is unlikely an
inelastic interaction could produce a high energy proton.

In conclusion, we have shown that elastic scattering cannot be
used to detect  a  bound $^{4}$n nucleus \cite{Ma02}. If the
$^{4}$n system exists it has to be very weakly bound and it will
also be a very large and diffuse system. It is very unlikely that
such a system can scatter protons at angles of
$\theta>90^{\circ}$.

We have benefited from useful discussions with N. Orr and F.M.
Marqu\'es. This work was supported by the National Science
Foundation under Grants No. PHY-007091 and PHY-00-70818.

\end{document}